\documentclass[preprint]{aastex}
\usepackage{emulateapj5}
\def\stacksymbols #1#2#3#4{\def\theguybelow{#2}
        \def\verticalposition{\lower#3pt}
        \def\spacingwithinsymbol{\baselineskip0pt\lineskip#4pt}
        \mathrel{\mathpalette\intermediary#1}}
\def\intermediary #1#2{\verticalposition\vbox{\spacingwithinsymbol
        \everycr={}\tabskip0pt
        \halign{$\mathsurround0pt#1\hfil##\hfil$\crcr#2\crcr
                \theguybelow\crcr}}}
\def\lta{\stacksymbols{<}{\sim}{2.5}{.2}}

\begin{document}

\title{CIRCULATION FLOWS: COOLING FLOWS WITH BUBBLE RETURN}

\author{William G. Mathews\altaffilmark{1}, 
Fabrizio Brighenti\altaffilmark{1}\altaffilmark{2},
David A. Buote\altaffilmark{3} \& Aaron D. Lewis\altaffilmark{3}}

\altaffiltext{1}{University of California Observatories/Lick Observatory,
Department of Astronomy and Astrophysics,
University of California, Santa Cruz, CA 95064, 
mathews@ucolick.org}

\altaffiltext{2}{Dipartimento di Astronomia,
Universit\`a di Bologna,
via Ranzani 1,
Bologna 40127, Italy, 
brighenti@bo.astro.it}

\altaffiltext{3}{University of California at Irvine, 
Dept. of Physics \& Astronomy, 4129 Frederick Reimes Hall,
Irvine, CA 92697, buote@uci.edu, lewisa@uci.edu}






\vskip .2in

\begin{abstract}
The failure of the XMM-Newton and Chandra X-ray telescopes to 
detect cooling gas in elliptical galaxies and clusters 
of galaxies has led many to adopt the position that 
the gas is not cooling at all and that heating 
by an active nucleus in the central E or cD galaxy 
is sufficient to offset radiative cooling. 
In this paper we explore an idealized limiting example 
of this point of view in which hot, buoyant bubbles formed 
near the center return the inflowing, radiatively cooling 
gas to distant regions in the flow.
We show that idealized steady state, centrally heated 
non-cooling flows can indeed be constructed.
In addition, the emission-weighted temperature profiles 
in these circulating flows resemble those 
of normal cooling flows.
However, these solutions are valid only (1) for a range 
of bubble parameters for which there is no independent 
justification, (2) for a limited spatial region in the cooling flow 
and (3) for a limited period of time after 
which cooling seems inevitable.
Our exploration of non-cooling flows is 
set in the context of galaxy/group flows.
%
\end{abstract}



\section{Introduction}

X-ray spectra of virialized hot gas associated with early type 
galaxies, galaxy groups and clusters 
indicate that the gas cools at less than a small fraction 
of the rate implied by their X-ray luminosities, 
(e.g. Peterson et al. 2001; B\"ohringer et al. 2002; 
Xu et al. 2002). 
There is no spectral evidence for gas at temperatures 
less than $\sim T_{virial}/3$.
As a result of these observations, 
it is often claimed that there is no cooling at all.
Many authors have speculated that the gas is heated 
by an active nucleus (AGN) in the cluster-centered 
E or cD galaxy and that this heating can explain the absence 
of gas at $T \lta T_{virial}/3$. 

The notion of AGN heating 
is also supported by the centrally disturbed gas visible 
in Chandra images of galaxy groups 
(e.g. Buote et al. 2003). 
Further evidence for hot bubbles on galaxy/group 
scales is provided by two (or limited multi) 
temperature spectral fits over an extended
central region of the X-ray emitting gas in galaxy groups 
(Buote 2002; 2003; Tamura et al. 2003).
The hot phase $T_h$ is typically 2 or 3 times hotter than the
cooler phase $T_c$, but $T_c(r)$ resembles the best-fitting
temperature profiles 
in single phase spectral models. 
On the larger scales of galaxy clusters 
that contain powerful radio sources, 
huge bubbles of hot gas are apparently formed 
by strong cluster-centered radio galaxies
and then buoyantly migrate out from the center of the flow.
(Fabian et al 2000; McNamara et al. 2000).
Although the absence of apparent cooling 
and the evidence for X-ray surface brightness fluctuations 
are shared by flows of all scales -- galaxies or clusters -- 
we are particularly interested here with  
galaxy/group flows because they may be less 
likely to have been disturbed by recent mergers 
and because the dominant stellar mass within 
the effective radius provides an independent measure 
of the gravitational potential.

The salient question that we address is:
Can central heating greatly reduce the
gas cooling rate ${\dot M}$ and also 
be consistent with the observed mean temperature of the hot gas?
To explore possible answers to this question,
we consider models in which bubbles of 
hot gas are heated to supervirial temperatures 
near the central AGN and then rise buoyantly 
to large radii. 
The bubbles move upstream in a normal cooling inflow 
in which gas slowly flows toward the central potential 
minimum, losing energy by radiative losses.  
To avoid cooling to intermediate and low temperatures, 
the heated bubbles must carry outward all the mass 
arriving at the center of the cooling flow.
The characteristic decline in the observed 
(single-phase) gas temperature
toward the center of these flows, $dT/dr > 0$, 
is rather clear 
evidence that normal radiative cooling is driving an inflow 
of gas toward the center. 
In the circulation flows we consider here, 
this cooling is assumed to be arrested by 
bubble formation associated with AGN heating. 

Flows in which cooling does not occur are expected to be 
in a steady state, at least on average. 
However, stellar mass loss and cooling or accretion inflow 
of hot gas from very large radii will cause the mass of gas 
in the circulation flow to increase slowly with 
time until some type of sudden cooling occurs. 
However, it is possible that idealized, 
quasi-steady circulation flows may persist for  
sustained periods without cooling, 
punctuated by intermittent cooling events.
Remarkably, there is little or no observational 
evidence of ongoing or recent global cooling 
events in well-observed flows.

Can heated gas moving upstream in the cooling flow 
carry away all the mass arriving at the center 
and still be consistent 
with the low central temperatures and positive temperature 
gradients determined by (single-phase) X-ray temperature 
observations?
This is a question of some interest to us since 
our studies of AGN-heated cooling flows using numerical 
gasdynamical simulations 
have been unsuccessful in agreeing with the 
central temperature profiles observed 
(Brighenti \& Mathews 2002; 2003).
When AGN heating is sufficient to arrest central cooling, 
the gas cools at larger radii in the flow and the
overall temperature gradient becomes negative throughout
a large central region, unlike the observations. 
However, more analytic and idealized solutions without numerical 
diffusion, such as we consider here, may capture 
details of the flow that cannot be easily duplicated 
in gasdynamical simulations. 
To keep the apparent temperature from rising sharply 
toward the central AGN,
it is necessary that the X-ray emission is not dominated by 
hot bubbles along the line of sight 
and that the inflowing gas is not heated too much 
by the counter-streaming bubbles.
We show here that idealized steady state circulation 
flows can be constructed that are consistent with 
observed temperature profiles, but only for bubbles 
in a restricted mass range and only for a limited 
spatial region and time.

The discovery of X-ray holes or bubbles in rich
clusters has resulted in a large number of numerical simulations.
Most of these simulations focus on 
the evolutionary interaction of 
powerful radio jets and lobes
from cluster-centered radio galaxies with the adjacent hot gas
and do not quantitatively 
address the essential question whether or
not strong radio jets can keep the gas from cooling to
low temperatures and still preserve the density and 
temperature profiles observed.
In the following section we briefly review some of these
numerical studies and contrast them with the the 
heated flows we consider here. 

\section{Recent Studies of AGN Heating and X-ray Holes}

The interaction of energetic supersonic jets with the hot gas 
has been simulated by many authors 
(e.g. Basson \& Alexander 2003; Reynolds, Heinz \& Begelman 2002).
These numerical studies may accurately portray the initial active 
phases in which a high velocity 
jet from a central AGN first enters the hot gas. 
But supersonic jets 
create strong bow shocks that heat the ambient gas within 
a large cocoon. 
In most X-ray observations 
there is no convincing evidence for 
extended shock-heated gas, 
although such phases may be relatively short-lived.

In response to the apparent absence of gas at 
supervirial temperatures, 
many jet-lobe simulations have been designed that do not produce
large masses of strongly shocked gas. 
These simulations typically involve
an {\it ad hoc} slow release of energy (and mass) 
in fixed regions at some distance from the central AGN 
(e.g. Br\"uggen 2003a; Br\"uggen \& Kaiser 2001, 2002;
Br\"uggen et al. 2002; Churazov et al. 2001;
Quilis, Bower \& Balogh 2001).
Br\"uggen and his collaborators have studied 
the consequences of injecting 
ultrahot gas with zero net velocity into fixed spherical regions 
of size $\sim 1$ kpc located $\sim 10$ kpc from the central AGN 
in galaxy clusters. 
Evidently, this type of energy release is 
an approximation to a stationary working surface of a jet. 
The hot gas atmosphere surrounding the AGN 
is initially in hydrostatic equilibrium 
with outward increasing entropy.
The luminosities of the energy injection, 
$L \sim 10^{42} - 10^{43}$ 
erg s$^{-1}$, are usually chosen 
for each simulation to be near the maximum possible $L$ 
that can still produce buoyant plumes and bubbles; larger $L$ 
would evacuate gas throughout the entire cluster 
core in times shorter than the buoyancy rise time 
(e.g. Br\"uggen 2003a). 
A common feature of these solutions is that cooler, low entropy 
ambient gas follows upward in a transient 
``entrained'' flow behind the rising 
hot gas. 
This transient, approximately 
axisymmetric convective circulation occurs as 
the rising gas is replaced by downflowing 
ambient gas surrounding the jet axis
that turns and flows upward behind the heated gas.
After $\sim 100$ Myrs, when the energy injection is 
turned off, some of this ``uplifted''  
low entropy gas typically begins to 
fall back toward the center, perhaps 
re-energizing the central AGN at a later time 
with a feedback mechanism.
While these papers are useful contributions to our 
understanding of the interactions of powerful radio 
jets with the hot gas, 
the authors often note that the cooling flow 
${\dot M}$ may be reduced if energy 
is released in this manner with duty cycle $\sim 0.5$.
However, the mass cooling rate 
${\dot M}(t)$ near the flow center is not explicitly calculated 
so it is difficult to link these simulations 
to the problem of the low 
upper limits on ${\dot M}$ set by XMM observations. 

By contrast, our numerical simulations study feedback 
heating and ${\dot M}$ reduction 
in genuine cooling flows over times comparable 
to the Hubble time
(Brighenti \& Mathews 2002, 2003).
The original gas is heated by the central AGN 
and no new mass is introduced.
${\dot M}(t)$ at the flow center and elsewhere is monitored 
throughout the entire calculation.
In our experience, 
when these cooling flows are sufficiently heated to 
reduce the central cooling rate ${\dot M}$, as XMM observations 
require, the azimuthally-averaged temperature and density 
profiles typically deviate from those observed. 
When cooling is suppressed by heating near the center, we find 
that it occurs at the same rate but at larger radii in the flow.
In flows with off-center jet-like heating,  
the gas at large angles relative to the heating axis 
easily flows toward the center and 
eventually cools to low temperatures.

The flow heating problem on galaxy/group scales 
that we consider here is very different from 
the heating of hot gas in richer galaxy clusters 
by powerful, linearly oriented radio jets and lobes.
Many flows with undetected ${\dot M}$ do not contain
powerful radio jets. 
For example, when observed with XMM-Newton, the galaxy/group 
NGC 5044 shows no RGS or EPIC 
spectral evidence of gas cooling to temperatures 
below about 0.7 keV 
(Tamura et al. 2003; Buote et al. 2003).
The (single phase) gas 
temperature profile drops from about 1.2 
keV at 30 kpc down to 0.7 keV at the origin, 
similar to many other E galaxies 
and qualitatively similar to $T(r)$ in rich clusters.
Also, like most elliptical galaxies, 
NGC 5044 has an unexceptional and unresolved 
weak central mJy GHz point source 
(NRAO/VLA Sky Survey; Condon et al. 1998). 
However, the {\it Chandra} image of NGC 5044 
(Fig. 1 of Buote et al. 2003) shows that the hot 
gas within at least $\sim 14$ kpc is highly disturbed 
in an azimuthally random and chaotic manner.
In addition, temperature fluctuations in the range
$0.7 \lta kT \lta 1.4$ keV are observed out to 
$\sim 30$ kpc (Buote et al. 2003).
The most important morphological attribute of 
these thermal and density 
irregularities is that they are 
symmetric about the center of NGC 5044,
with no evidence of jet-like orientation. 
If AGN heating is the agency that keeps the gas from 
cooling, the heating 
in NGC 5044 and other similar E galaxies must 
be more radially symmetric than 
the jet-heating considered 
in the radio galaxy simulations discussed above. 
Alternatively, because of the short dynamical times
in galaxy scale flows, the jet axis in would have to be 
randomly reoriented every $\sim 10^6$ years 
to maintain the appearance of symmetric heating. 
Finally, 
the outward mass flux ${\dot M}$ is likely to be maximized 
in spherically symmetric bubble outflows 
that move away from the central AGN in every direction. 

Ruszkowski \& Begelman (2002) and Br\"uggen (2003b) 
have studied non-cooling, 
static models of cluster gas atmospheres. 
The gas is heated at large radii 
by thermal conduction and at small radii by a cloud of tiny 
expanding bubbles, a heating mechanism 
proposed by Begelman (2001). 
The heating is spherically symmetric, 
similar to the flows we discuss here. 
These cluster-scale 
models are almost uniquely successful in preserving 
the positive temperature gradient observed. 
However, this combination of central heating and conduction 
does not work for flows on galaxy/group scales; 
even with the maximum (Spitzer) conductivity 
it is impossible to get both small ${\dot M}$ and 
acceptable temperature gradients 
(Brighenti \& Mathews 2003).
In the heating scheme proposed by Ruszkowski \& Begelman 
a large number of very small ``effervescent'' bubbles are
produced near the central AGN and buoyantly flow into 
the hot gas throughout a large volume, 
acquiring a steady state radial distribution.
The bubbles do not mix thermally with the ambient gas 
but expand adiabatically,
heating the ambient flow by $PdV$ work as they expand.
While this bubble model has some features in common 
with that considered here, 
we find that very small bubbles are inconsistent 
with steady state circulating flows. 
%

In the following discussion we investigate the possibility that 
hot gas in galaxy/groups similar to NGC 4472 or 
NGC 5044 is heated near the center of the flow 
and converted into hot bubbles. 
The rising bubbles are required to carry mass outward 
at the same rate that mass arrives in the core 
by normal radiative losses in the interbubble flow.
The interbubble flow receives both momentum and 
energy from the rising bubbles. 
To optimize the outward rate of mass transfer, 
the heating is assumed to be symmetric about the 
central AGN, not confined to a single jet.

\section{Circulation Flows}

We explore here the conditions required 
to establish oppositely moving, two phase, steady state flows 
in galaxy/groups 
in which the gas does not cool to low temperatures.
A cool phase near the virial temperature 
flows inward as in a normal cooling flow.
At some small radius $r_h$ we assume this flow is 
heated and transformed 
into hot bubbles that buoyantly float upstream in the 
cooling flow, returning the mass flux received in the incoming 
flow to some large circulation radius $r_c$
where the bubble entropy matches
that of the local cooling flow gas. 
Gas circulates in the gravitational field 
like a thermodynamic engine.
The mechanism that produces the bubbles is not 
specified and for simplicity the physical condition of the gas 
within $r_h$ or beyond $r_c$ are not considered.
Clearly, 
an idealized steady flow model like this 
cannot be strictly correct since the mass 
of circulating gas is expected to gradually increase due to 
stellar mass loss and inflow of gas from $r > r_c$.
As the mass of circulating gas slowly increases, the 
increasing gas density should eventually lead to 
extensive cooling. 
Nevertheless, a quasi-steady circulation model could
be a reasonable representation for a limited
time between cooling episodes.
Unlike the numerical simulations described earlier,
no new mass is injected into the flow.

We also assume that 
the hot bubbles retain their coherency as they move 
through $\sim 10 - 15$ bubble diameters.
This degree of coherence 
is in fact observed in the Perseus Cluster 
(Fabian et al. 2000), assuming that the distant ``ghost cavities'' 
were formed closer to the central AGN.
The hot bubbles simulated by 
Churazov et al. (2001) 
float upward in the cooling flow atmosphere
and come to rest at some large radius where the ambient entropy
is equal to that within the bubble.
However, we have found 
bubble evolution to be a difficult problem for numerical
gasdynamics and  
we remain sceptical of our assumption of bubble coherency.
The bubbles in Perseus may be stabilized  
by internal magnetic fields. 
It would be relatively easy to modify our 
circulation model to include 
bubble fragmentation according to some {\it ad hoc} prescription,
but the effect of bubble fragmentation can easily be imagined by 
comparing similar circulation flows having different constant 
bubble masses.

In our circulating flows the hot bubbles do not thermally mix
(by thermal conduction for example) with the incoming flow 
before reaching the circulation radius $r_c$ 
where the bubble and inflow entropies are approximately equal. 
Thermal mixing is another effect that could be included in 
more complicated circulation flows, 
but our assumption of no thermal mixing is a limiting 
case of interest.
If premature bubble-flow thermalization occurs, 
the inflowing gas will be heated and the apparent temperature 
profile is likely to become negative, unlike the  
(single phase) temperature profiles observed.
Unintended thermal mixing may have occurred in our numerical 
simulations of heated cooling flows because of the limited 
numerical resolution
(Brighenti \& Mathews 2002; 2003). 
When the bubbles and interbubble flow are viewed 
along the same line of sight, 
an apparent positive $dT/dr$
can be maintained if the filling factor 
$f_b$ of the rapidly moving bubbles is relatively small 
and if the bubbles do not significantly heat the ambient gas. 
We show that 
the constraint $f_b < 1$ also limits the 
heating that bubbles can receive at 
the heating radius $r_h$ and therefore 
also the radial extent $r_c$ of the circulation.

Finally, to keep our circulation models as simple as 
possible, we have resisted a temptation to include 
an additional pressure component within the bubbles 
due to (relativistic) cosmic rays or magnetic fields. 
An additional nonthermal pressure could help 
increase the outer radius $r_c$ of the circulation, 
but may transport less mass outward.
Nevertheless, many of the same constraints we derive 
here for purely thermal circulations also apply 
to bubbles with nonthermal buoyancy.

Our approximate, idealized circulation flows 
are not intended to reproduce the observed properties of 
any particular group or cluster although we use 
parameters for the massive elliptical galaxy NGC 4472 
whenever appropriate. 

\clearpage

\section{Inflow Equations}

The steady state equation of continuity for the inflowing gas 
\begin{equation}
{\dot M} = 4 \pi r^2 f \rho u
\end{equation}
contains the fraction of the 
volume $f(r)$ filled by the inflowing gas. 
The bubble filling factor is $f_b = 1 - f$. 
Both ${\dot M}$ and $u$ are negative.
The equation of motion and the conservation of thermal 
energy in the cooling inflow are
\begin{equation}
\rho u { d u \over dr } = - {d (\rho \theta)  \over dr}
- \rho g_e
\end{equation}
and
\begin{equation}
\rho u {3 \over 2} { d \theta \over dr} = 
\theta u {d \rho \over dr} - { \rho^2 \Lambda \over m_p^2}.
\end{equation}
Here we define $\theta = k T/ \mu m_p$ 
where $\mu$ is the molecular weight
so the gas pressure is $P = \rho \theta$. 
The radiative cooling coefficient $\Lambda(T,z)$ 
erg cm$^3$ s$^{-1}$ is based on the results of 
Sutherland \& Dopita (1993). 
The effective gravity $g_e$ includes the momentum 
imparted to the inflowing gas by bubbles moving in 
the opposite direction (see \S 5 below).

These equations can be written as two coupled 
differential equations,
\begin{equation}
{du \over dr} = {u \over r}
{[ 5 \alpha \theta - 3 g_e + (2 r \rho \Lambda / m_p^2 u)] 
\over (3 u^2 - 5 \theta)}
\end{equation}
and 
\begin{equation}
{d \theta \over dr} = {2 \over u r}
{[ u \theta (g_e r - u^2 \alpha) 
+ (r \rho (\Lambda / m_p^2) (\theta - u^2) ]
\over (3 u^2 - 5 \theta)}
\end{equation}
and $\rho(r)$ is found from Equation (1).
In deriving these equations, 
$d \rho / dr$ has been eliminated by differentiating 
Equation (1) and the variation of the filling factor 
is included in the variable 
\begin{equation}
\alpha  = 2 + { d \log f \over d \log r}.
\end{equation}

The gravitational acceleration $g$ is based on the massive
E galaxy NGC 4472 which is embedded in a galaxy group having
an NFW dark matter halo.
The stellar mass
has a de Vaucouleurs profile
(total mass: $M_{*t} = 7.26 \times 10^{11}$ $M_{\odot}$;
effective radius: $R_e = 1.733' = 8.57$ kpc)
with a core
$\rho_{*,core}(r) = \rho_{*,deV}(r_b)(r/r_b)^{-0.90}$
within the break radius $r_b = 2.41'' = 200$ pc
(Gebhardt 1996; Faber et al. 1997).
The NFW halo has total mass $M_h = 4 \times 10^{13}$ $M_{\odot}$.
For an excellent fit to the gravitational force in this
stellar-dark matter potential we use
\begin{equation}
g = \left[ (a r_{kpc}^{p} )^{-s}
+ (b r_{kpc}^{q})^{-s} \right]^{-1/s}~~~{\rm cm~s}^{-2}
\end{equation}
with $a = 1.0233 \times 10^{-6}$, $p = 0.1$,
$b = 4.0378 \times 10^{-7}$, $q = -0.8354$ and $s = 3$.

The hot gas density observed in NGC 4472 is found 
by combining data from {\it Einstein} (Trinchieri, Fabbiano,
\& Canizares 1986) and
ROSAT (Irwin \& Sarazin 1996).
For the inner region
we have Abel-inverted Chandra surface brightness data from
Loewenstein et al. (2001)
and normalized them to previous observations.
These data can be accurately fit with 
\begin{equation}
n_e(r) =  \sum_{i=1}^{4} { n(i) \over 
\{1 + [r /R_e r(i)]^{p(i)}\} }
\end{equation}
where 
\begin{eqnarray*}
n(i) & = & (0.117, 0.00735, -0.000493, 0.142)\\
r(i)/R_e & = & (0.107, 0.95, 10., 0.04) \\
p(i) & = & (2.0, 1.14, 1.19, 3.0).
\end{eqnarray*}
The gas temperature in NGC 4472 
\begin{equation}
T(r) = 1.55 \times 10^7 { {r_{kpc} + 2.7} \over {r_{kpc} + 5.98}}
\end{equation}
is a fit to the combined {\it ROSAT} HRI and PSPC
observations of Irwin \& Sarazin (1996)
and {\it ROSAT} PSPC observations of Buote (2000).

\subsection{Flow Without Bubbles}

The cooling flow equations are solved indirectly by requiring 
that the gas density match the observed density in 
NGC 4472 at the end points of the circulation, 
$r_h$ and $r_c$.
To illustrate a typical solution of this two-point boundary 
value problem,
we consider a circulation flow between radii $r_h = 1$ and 
$r_c = 10$ kpc.
The instantaneous rate that gas cools in this annular region 
can be determined from 
\begin{equation}
{\dot M} \approx \int_{r_h}^{r_c} {\rho^2 \Lambda \over m_p^2}
{2 \mu m_p \over 3 k T} 4 \pi r^2 dr \approx 0.7~~~
M_{\odot}~{\rm yr}^{-1}
\end{equation}
using the observed $T(r)$ and $\rho(r)$ profiles  
and evaluating $\Lambda$ with an assumed abundance 0.5 solar. 

Figure 1 shows a typical 
solution of Equations (1), (4) and (5) 
for a cooling flow without bubbles ($f = 1$, $\alpha = 2$, 
and $g_e = g$) between $r_1$ and $r_2$ assuming 
${\dot M} = 0.7$ $M_{\odot}$ yr$^{-1}$.
When the flow equations are solved iteratively in the 
manner described, both the temperature profile and its end-point 
values are completely determined by 
$\rho(r_h)$ and $\rho(r_c)$.
The modest disagreements apparent in Figure 1 between the 
computed temperature and density and the profiles observed 
in NGC 4472 are expected because of the approximations 
we have made: steady state flow, suppression of mass loss from 
stars, no gas extending beyond $r_c$, etc.
When the flow is subsonic ($u^2 \ll \theta$)  
the temperature gradient $d \theta/dr$ (Equation 5) has 
the same sign as $u g_e + \rho \Lambda /m_p^2$.
Since $u < 0$, it is clear that the radiative term is responsible 
for the positive temperature gradients observed.
Although the flow temperature gradient in Figure 1 
is flat, $dT/dr \approx 0$, 
this solution is entirely adequate for our 
purposes here: to determine the influence 
of counter-moving bubbles on the observed thermal properties 
of the flow. 
We are primarily interested in {\it differential} changes in the 
temperature profile when bubbles are introduced.

\section{Bubble Dynamics}

A circulation flow without cooling below $T(r_h)$ 
can be maintained if the 
inflowing gas is heated at radius $r_h$, 
forming bubbles of hot gas that rise 
buoyantly upstream in the cooling flow. 
At $r_h$ the bubbles have temperature $T_b(r_h) = hT(r_h)$ where
$T(r_h)$ is the local cooling flow temperature and $h > 1$ is 
a dimensionless heating factor.
To avoid needless complications, 
all bubbles are assumed to have the same mass
$m_b = \rho_b (4/3) \pi r_b^3$ 
for each solution; the consequences of bubble breakup can be 
inferred from solutions with different $m_b$.
As they move outwards, the hot bubbles evolve adiabatically 
($\gamma = 5/3$) in pressure equilibrium with the ambient gas.
The density inside the bubbles is 
\begin{equation}
\rho_b = { \rho^{3/5} \rho_h^{2/5} \over h}
\left( {\theta \over \theta_h} \right)^{3/5}
\end{equation}
where $\rho_h = \rho(r_h)$ and $\theta_h = \theta(r_h)$.

The equation of motion of the bubbles is dominated by the drag 
force so the terminal velocity is reached almost immediately 
after they are formed.
When moving 
at the terminal velocity, the forces on the bubble balance:
\begin{equation}
m_b g + {\langle {d P \over dr} \rangle} {4 r_b \over 3} \pi r_b^2 + 
\delta \rho_f (u_b - u)^2 \pi r_b^2 = 0
\end{equation}
where $u_b$ is the bubble velocity. 
The external pressure force on the bubble is 
the pressure gradient 
across the bubble $dP /dr$ 
times the mean chord through a sphere 
$4 r_b /3$ and the bubble area $\pi r_b^2$.
The final term in Equation (12) is the approximate 
drag force on the bubble where $\rho$ is the density in 
the ambient cooling flow 
and $\delta$ is a dimensionless coefficient 
of order unity that depends on the shape of the 
bubble and other imponderables; we assume $\delta = 0.5$.
Since the cooling flow is highly subsonic, 
hydrostatic equilibrium is a good approximation, 
$dP/dr \approx -\rho g$, 
and the bubble velocity follows from Archimedes' principle, 
\begin{equation}
u_b = u + \left[ { g \over \delta \beta }
\left( {1 \over \rho_b} - {1 \over \rho} \right) \right]^{1/2}
\end{equation}
where
\begin{equation}
\beta = {\pi r_b^2 \over m_b} = {\pi \over m_b}
\left( { 3 m_b \over 4 \pi \rho_b } \right)^{2/3}.
\end{equation}

The collective drag interactions of a multitude of rising bubbles 
exerts an outward force on the cooling flow so the 
effective gravity becomes
\begin{equation}
g_e = g - \delta (u_b - u)^2 (1 - f) \rho_b \beta.
\end{equation}
Typically, $g_e$ differs from $g$ by $\lta 5$ percent.

The energy transmitted from the bubbles to the surrounding 
gas is of particular interest for 
the cooling flow problem since the observed 
temperatures are always lowest at the center
(as in Equation 9).
This observation indicates that most of the 
X-ray emitting gas near the flow center 
is not being strongly heated 
even when heating is required to account for the 
absence of further cooling. 

The nature of the energy exchange between the 
bubbles and the flow is not well understood.
Several authors 
(Br\"uggen \& Kaiser 2002; Churazov et al. 2002
Begelman 2001; Br\"uggen 2003b)
have suggested that 
cooling flows are heated by
the $PdV$ work expended as bubbles expand during 
their ascent in the flow. 
While some $PdV$ heating by bubbles 
may occur during the transient adjustment when bubbles 
first move into a region of the flow devoid of bubbles, 
it is less obvious that strong $PdV$ heating is 
expected in steady state flows where the filling 
factors of the bubbles and flow 
remain constant at every radius, i.e. $dV = 0$. 
In steady flow as expanding bubbles rise subsonically upward 
into a spherical shell of interbubble gas, 
the volume of interbubble gas in that shell 
has already been reduced to accommodate the expanded bubbles;
this volume adjustment is contained in the filling factor $f(r)$ 
for that shell. 
Since the bubbles move subsonically, 
the flow in the interbubble gas as it is rearranged at  
constant volume in the shell is also subsonic and therefore largely 
incompressible, i.e., $dV = 0$.
$PdV$ heating is likely to depend on the filling 
factor in some complicated manner that we do not 
investigate here. 
In spite of these reservations, we consider 
the possibility that $PdV$ heating may be important. 

In steady state a single rising bubble of increasing 
volume $V_b = m_b /\rho_b$ delivers $PdV$ power at a rate
$$
{\dot E}_{pdv} = P_b {d V_b \over dt} = 
P_b (u_b - u) {d V_b \over dr} 
$$
$$
~~~~~~~~
= -P_b (u_b - u) {m_b \over \rho_b r} 
{r \over \rho_b} { d \rho_b \over dr}
= (u_b - u) {m_b \rho g \over \gamma \rho_b}
~~~{\rm erg}~{\rm s}^{-1}
$$
where we assume $P = P_b \propto \rho_b^{\gamma}$
and $dP / dr = - \rho g$.
Multiplying by the space density of bubbles,
$N_b = (1 - f) \rho_b / m_b$, we find the 
power delivered to the inflowing gas by all bubbles,
$$
{\dot \varepsilon}_{pdv} = e_{pdv} N_b {\dot E}_{pdv} =
{e_{pdv} \over \gamma} (1 - f) (u_b - u) \rho g,
~~~{\rm erg}~{\rm cm}^{-3}~{\rm s}^{-1}
$$
where $e_{pdv}$ is the uncertain efficiency that 
the expanding bubbles heat the gas locally. 

Regardless of the uncertain relevance of $PdV$ heating,
the cooling inflow is certainly heated by 
the drag interaction with the counterstreaming bubbles. 
The drag-related power generated by a single bubble is 
the product of the drag force and the relative velocity
$F_d (u_b - u) = \delta \rho (u_b - u)^3 \pi r_b^2$.
The local density of bubbles is $N_b = \rho_b (1-f)/m_b$
so the collective power density produced by all bubbles 
is $F_d (u_b - u) N_b$.
The bubbles could heat the surrounding gas by producing 
(weak) shocks or turbulence that dissipates locally. 
Non-local heating could result if the energy received 
from bubbles was in the form of sound waves, 
gravity waves or long-lived turbulence.
Since the fraction of this energy that goes into local  
heating is unknown, we include a heating efficiency 
factor $e_d$ in the power density. 
The heating efficiency of rising bubbles is discussed 
by Churazov et al. (2002).
Bubble-flow drag interactions heat the gas at a rate
$$
{\dot \varepsilon}_{d} = { 3 e_d \over 4 r_b}
\delta (1 - f) \rho (u_b - u)^3
~~~{\rm erg}~{\rm cm}^{-3}~{\rm s}^{-1}.
$$
 
Bubble heating can be included in the flow equations 
by replacing the radiative cooling coefficient $\Lambda$ 
in Equations (4) and (5) with 
$\Lambda + \Lambda_{pdv} + \Lambda_d$ where 
\begin{equation}
\Lambda_{pdv} = 
- {3 \over 5} e_{pdv} (1 - f) m_p^2 
{g (u_b - u) \over \rho }
\nonumber 
~~~{\rm erg}~{\rm cm}^{3}~{\rm s}^{-1}
\end{equation}
and 
\begin{equation}
\Lambda_d = 
- {3 \over 4} e_d \delta (1 - f) {m_p^2 \over \rho}
{(u_b - u)^3 \over r_b} \nonumber
~~~{\rm erg}~{\rm cm}^{3}~{\rm s}^{-1}.
\end{equation}

\subsection{Steady State Circulation Flows}

The essential and 
necessary condition for steady state circulation
requires that the rate of cooling mass inflow ${\dot M}$ 
is balanced at every radius 
by a proportional upstream mass flow 
${\dot M}$ carried by the bubbles, 
\begin{equation}
{\dot M} = 
4 \pi r^2 f \rho u = - 4 \pi r^2 (1 - f) \rho_b u_b
\end{equation}
so that
\begin{equation}
f = { \rho_b u_b \over \rho_b u_b - \rho u }.
\end{equation}

The luminosity expended by AGN heating at radius $r_h$,
$$
L_{agn} = h {3 k T(r_h) \over 2 \mu m_p} {\dot M}
$$
$$
~~~~~~~~~~~~~~~~~
\approx 1.3 \times 10^{41}
h \left( { T \over 10^7~{\rm K}} \right)
\left( { {\dot M} \over M_{\odot}~{\rm yr}^{-1}} \right)
~~~{\rm erg}~{\rm s}^{-1},
$$ 
can be produced by accretion onto the central black hole 
with mass flow ${\dot M}_{bh} = L_{agn}/\eta c^2$
where $\eta \approx 0.1$ is the 
efficiency that the accreted mass converts to energy.
Combining these expressions,
$$
{\dot M}_{bh} \approx 2 \times 10^{-5} 
h \left( { T \over 10^7~{\rm K}} \right)
\left( { {\dot M} \over M_{\odot}~{\rm yr}^{-1}} \right)
~~~{\rm M}_{\odot}~{\rm yr}^{-1}.
$$
Since ${\dot M}_{bh} \ll \ {\dot M}$, 
only a negligible flow into the hole can provide the 
necessary heating at $r_h$ so 
the mass flux in both radial directions 
in the flow beyond $r_h$ must exactly balance at every radius 
(Equation 18).

Counterstreaming flow solutions are found 
by iteratively 
solving Equations (1), (4), (5), (11), (13) and (19). 
We begin with a bubble-free flow as in Figure 1, 
then calculate 
the dynamics of the bubbles required to 
carry the same ${\dot M}$. 
The total volume of the bubbles at each radius 
provides an estimate of 
the filling factor $f(r)$, 
which can then used to update the (iterative) solution 
for the inflowing gas.
The iteration of the two point boundary problem for the 
cooling inflow (Equations 4 and 5) is 
embedded within the global iterative scheme involving 
the bubble dynamics.

For simplicity we fix the inner radius $r_h$ of the 
circulation for all calculations, 
but the outer radius $r_c$ is free to vary as required 
by the solution. 
With each bubble-flow 
iteration, the radius of the circulation flow  
$r_c$ is determined as that radius where 
the entropy of the rising bubbles equals that in 
the cooling inflow;
the cooling flow density is then renormalized to the
local density of NGC 4472 at $r_c$.
At $r = r_c$ the density and 
temperature inside the bubbles naturally become the same 
as those in the local flow, 
and the bubbles merge with the cooling flow. 
The outward velocity of the bubbles at $r_c$ 
and the inward velocity of the cooling flow gas 
are both very small and highly subsonic 
at $r_c$, but never exactly zero;
we assume that this small differential velocity 
is dissipated without appreciably altering the 
local entropy.
As before, the temperature profile of the interbubble flow 
and its end-point values $T(r_h)$ and $T(r_c)$ 
are completely determined by the 
requirement that the gas density match that 
observed in NGC 4472 at $r_h$ and $r_c$.

For a rigorously consistent solution the global mass 
flow rate ${\dot M}$ should be redetermined from 
Equation (10) (using the computed flow) 
for each value of $r_c$ as it changes 
during the iterative calculation.
However, we have found that this requirement is 
not essential for overall convergence and 
for simplicity we have 
decided to relax this additional constraint,  
keeping both $r_h = 1$ kpc and 
${\dot M} = 0.7$ $M_{\odot}$ yr$^{-1}$ constant 
for all solutions.  
For given ${\dot M}$ and $r_h$, 
the important parameters for each circulation 
flow are the heating factor $h$, 
the bubble mass $m_b$ and the heating efficiencies  
$e_{pdv}$ and $e_d$ of the bubbles.

\section{Restrictions on Flow Parameters}

Not every set of flow 
parameters -- $h$, $m_b$, $e_d$, $e_{pdv}$, ${\dot M}$ and $r_h$ -- 
results in a solution that converges. 
Successful solutions must satisfy 
several additional conditions: 
(1) the bubble radius must not exceed the local radius in the 
flow,
(2) the bubbles must not fill the entire volume 
and choke off the inflowing gas, and 
(3) the mass of the bubbles must be sufficient to carry the
outward mass flux ${\dot M}$.

The first obvious restriction is most acute near 
the heating radius $r_h$ where $r_b/r$ is largest.
The requirement that 
the size of the bubbles not exceed $r_h$ is  
\begin{displaymath}
r_h > r_b = 
0.44 \left( { m_b \over (10^6~M_{\odot})}
{h \over (n_e(r_h) / 0.1~{\rm cm}^{-3})} \right)^{1/3}
~~~{\rm kpc}. 
\end{displaymath}
The second condition is that the filling factor 
for the bubbles $f_b = 1 -f$ cannot exceed unity near $r_h$ 
where the bubble density and flow congestion is highest. 
If $t_{bd} = 2 r_b/ [u_b(r_h) - u(r_h)]$ 
is the time for bubbles to move 
one bubble diameter at $r = r_h$, 
$t_b = m_b/{\dot M}$ is the time interval between bubble 
formation, 
$V_b = (4/3) \pi r_b^3$ is the bubble volume, 
and $V_h \approx 8\pi r_h^2 r_b$ is the volume of an annular 
shell of thickness $2r_b \ll r_h$, then we require that 
\begin{equation}
f_b(r_h) = {t_{bd} V_b \over t_b V_h} < 1.
\end{equation}
This condition can be written
\begin{equation}
{ {\dot M} \over (M_{\odot}~{\rm yr}^{-1}) }
{ h \over {\cal M}_{bh} T_7^{1/2} r_{h,kpc}^2 
(n_h/0.1~{\rm cm}^{-3}) } \lta 17
\end{equation}
where 
${\cal M}_{bh} = [u_b(r_h) - u(r_h)]/(5 \theta_h /3)^{1/2}$ 
is the Mach number of the relative flow 
and $n_h = (\rho_h /m_p)(2 + \mu)/(5 \mu)$.
Condition (21) sets an upper bound on the amount of heating 
$h$ that can occur at $r_h$. 

Finally, from Equation (18) the mass flux carried by 
the bubbles is 
${\dot M} = 4 \pi r^2 (1-f) \rho_b u_b$.
Near the heating radius, $r \approx r_h$, where 
$\rho_b < \rho$ and $u_b > |u|$, 
the bubble velocity 
$u_b \approx (g/\delta \beta \rho_b)^{1/2}$. 
Therefore, 
\begin{equation}
{\dot M} \approx 4 \pi r_h^2  [1 - f(r_h)]
\left( {4 \pi \over 3} \right)^{1/3}
\left( {\rho_h \over h} \right)^{5/6}
\left( {g \over \pi \delta} \right)^{1/2}
m_b^{1/6}
\end{equation}
where $\rho_b \approx \rho_h/h$. 
For a given cooling flow where ${\dot M}$, 
$r_h$, $\rho_h$, $f$ and $g$ are all specified or 
constrained, the bubble mass $m_b$ must be 
large enough to carry 
the mass flow ${\dot M}$ dictated by the cooling inflow.
Although the dependence on $m_b$ 
is weak, $m_b$ and $h$ are the only flexible parameters. 
As $m_b$ decreases, the central heating 
factor $h \propto m_b^{1/5}$ must also decrease 
for fixed ${\dot M}$.
If bubbles of very small mass $m_b$ are formed at $r_h$ or are
produced later on by the fragmentation of larger bubbles,
condition (22) can cause 
the maximum allowed heating $h$ to fall below
unity for which no circulation flows are possible. 
\section{Results}

\subsection{Flows Without Bubble-Flow Energy Exchange}

Results for a sample of steady circulation flows are 
illustrated in Figure 2.
The heating efficiencies $e_d$ and $e_{pdv}$ that the flow 
is heated by the bubbles are 
assumed to be zero and the mass flow rate  
${\dot M} = 0.7$ $M_{\odot}$ yr$^{-1}$ 
and heating radius $r_h = 1$ kpc are constant 
for all flows.
Two circulation flows are 
shown in Figure 2 for each assumed bubble mass $m_b$, 
one with a typical value $h$ and one near the largest
$h = h_{max}$ for which solutions are physically possible 
or appropriate.
Each solution in Figure 2 shows the cooling inflow 
variables $n_e(r)$, $T(r)$ and $u(r)$ with solid curves. 
In the density panels the solid curves begin and end 
on the observed density profile for NGC 4472, 
which is shown with dotted lines in the top row of panels.
When this flow reaches $r_h = 1$ kpc, the gas is heated
by a factor $h$
and the density and temperature change abruptly to 
bubble values, 
$\rho_b(r_h) = \rho(r_h) /h$ and $T_b(r_h) = h T(r_h)$ 
respectively.
Then the bubbles rise in the flow along the dashed 
lines until they eventually intersect the cooling inflow 
at some large radius $r_c$.
The entropy of the bubbles and flow are identical at $r_c$ 
and the bubbles provide just enough gas to maintain 
the inflowing cooling flow from that radius. 
The third row of panels 
({\it c}, {\it g} and {\it k}) shows the negative 
cooling inflow with solid lines and the much faster
outward motion of the rising bubbles, 
$u_b > 0$, with dashed lines. 
The variation of the bubble radius $r_b(r)$ is shown 
in the bottom row of panels 
({\it d}, {\it h} and {\it l}) with dash-dotted lines.
The dashed lines in these bottom panels 
show the variation of the inflow volume filling factor $f(r)$.
As $h$ approaches the highest possible value for each 
bubble mass $m_b$, $h \rightarrow h_{max}$, 
the bubbles nearly fill the entire 
available volume at $r_h$ where $f \rightarrow 0$, 
as explained by Condition (21).

For the most massive bubbles considered,
$m_b = 10^6$ $M_{\odot}$, 
solutions for two heating parameters are shown, 
$h = 3$ and 6 for which the circulation radii 
are $r_c = 10$ and 25 kpc respectively.
The density, temperature and velocity 
profiles for the cooling inflows 
for these two solutions are almost identical.
We have not considered larger values of 
$h$ because the bubble velocity $u_b$ for the $h = 6$ solution 
approaches the sound speed in the cooling flow
$(5 \theta /3)^{1/2} = 476~(T/10^7~{\rm K})$ km s$^{-1}$.
Larger $h$ would result in supersonic 
bubble velocities that would drive shocks into the cooling gas, 
dramatically increasing its temperature.
Also the bubble size at $r_h$ for the $h = 6$ solution,
$r_b(r_h) \sim 1$ kpc is comparable to $r_h$, 
again arguing against larger values of $h$.

The central column of Figure 2 illustrates 
two circulation flows for 
bubbles of mass $m_b = 10^5$ $M_{\odot}$ 
for $h = 3$ and 6.5.
The circulation with $h = 6.5$ extends out to 
$r_c = 18$ kpc,
but the flow filling factor $f$ is less than about 0.1
at $r_h$ so there are no circulations with 
$h > 6.5 \approx h_{max}$ (Condition 21).
In order for this marginal flow 
to support mass flux ${\dot M}$, the relative flow 
$u_b - u$ must increase as $r \rightarrow r_h$
and the drag increases accordingly.
The bubble velocity $u_b$ is seen to decline toward $r_h$.
As a result of the large drag, 
the effective gravity $g_e$ 
(Equation 15) is directed outward 
near $r_h = 1$ kpc where the flow would be 
Rayleigh-Taylor unstable.
This region of instability becomes larger for 
higher values of $h$. 
Because of the approximations we have made, 
the density in the flow region $r_h < r < r_c$ is somewhat  
higher than the density observed in NGC 4472 (dotted lines).
However, if this flow were observed, the apparent density 
$n_{e,app}$ would be lowered since the flow only occupies 
a fraction $f$ of the volume, 
i.e., $n_{e,app}(r) \approx f^{1/2} n_e(r)$.

The first column in Figure 2 shows the circulation pattern 
for bubbles of mass $m_b = 3 \times 10^4$ $M_{\odot}$ 
for two flows with $h = 3$ and 3.6.
The flow filling factor $f$ is dropping rapidly 
as $r \rightarrow r_h$ for the $h = 3.6$ solution so this 
is close to the maximum possible heating (and $r_c$) for 
these smaller bubbles.
We were unable to achieve fully 
convergent flow solutions with even smaller 
bubbles of mass $m_b \lta 10^4$ $M_{\odot}$. 
While our search for such solutions was not exhaustive, 
it is apparent from the trend in panels 
{\it k} $\rightarrow$ {\it g} $\rightarrow$ {\it c} in Figure 2 
that the maximum values of $u_b$, $h$ 
and $r_c$ all decrease with $m_b$, 
as expected from condition (22).
Bubbles of mass $m_b = 3 \times 10^4$ $M_{\odot}$ 
are approaching the minimum $m_b$ necessary to 
carry the cooling mass flow ${\dot M}$ for any 
heating $h > 1$.
Non-circulating 
flows with bubbles of lower mass must ultimately 
cool radiatively to temperatures much lower than 
$T(r_h)$, in disagreement with XMM observations.

One of the motivations for our exploration of 
non-cooling circulation flows is to estimate the 
influence of the bubbles on the apparent gas temperature 
which typically has a minimum near the origin.
To simulate the apparent temperature that would be inferred 
when viewing the cooling inflow and the hot rising 
bubbles along the same line of sight,
we calculated the local emission-weighted temperature
\begin{equation}
\langle T\rangle = { \rho^2 T f + \rho_b^2 T_b (1-f) \over
\rho^2 f + \rho_b^2 (1-f) }
\end{equation}
which is plotted as dotted lines for each of the flows 
in panels {\it b}, {\it f} and {\it j} of Figure 2.
For most circulation flows $\langle T\rangle$ 
is nearly indistinguishable from the temperature of the
cooling flow component.
Consequently, a single phase interpretation of the
temperature profile would show a mildly positive gradient,
$dT/dr > 0$, similar to those observed and consistent 
with traditional cooling flows. 
This is a desirable feature of the circulation model.
Only for the least massive bubbles considered,
$m_b = 3 \times 10^4$ $M_{\odot}$, does
$\langle T\rangle$ rise slightly as $r \rightarrow r_h$,
but the differential change in the temperature 
profile caused by the hot bubbles is small.
The apparent density $\rho_{app} =
(\langle \rho^2 \rangle)^{1/2}$ and entropy profiles are also
similar to those of the cooling flow component alone.
We stress that these results differ from normal convection
where (1) there is a single temperature and density
at every radius (apart from fluctuations),
(2) the temperature gradient
is determined by the gravitational potential,
$dT/dr = -(2 \mu m_p/5 k)(d \Phi /dr)$,
and therefore has a sign opposite to that observed,
and (3) the entropy is constant.

\subsection{Flows With Bubble-Flow Energy Exchange}

The ratio of $PdV$ heating to drag heating is
$$
{ {\dot \varepsilon}_{pdv} \over {\dot \varepsilon}_{d} }
= {4 \over 5} {e_{pdv} \over e_d}
{g r_b \over \delta (u_b - u)^2 }
\approx 1.0 { (r_b / 100~{\rm pc}) \over
\delta ~ r_{kpc}^{0.85} (u_b / 100~{\rm km/sec})^2 }
$$
where $u_b \ll |u|$ and 
we use $g \approx 4.47 \times 10^{-7} r_{kpc}^{-0.85}$ 
cm s$^{-2}$ which is appropriate 
for $1 \lta r_{kpc} \lta 100$ in NGC 4472.
The result ${\dot \varepsilon}_{pdv}/{\dot \varepsilon}_d 
\sim 1$ follows immediately from 
Equation (13) provided $u_b \ll |u|$.
In view of the uncertain physical nature of these two 
comparable bubble-flow heating mechanisms, we consider 
heating due to bubble expansion and drag separately. 

Figure 3 illustrates four representative cooling flow models 
in which the work done by the drag force is assumed to 
heat the local cooling flow gas with various 
efficiencies $e_d$ (and $e_{pdv} = 0$).
As before ${\dot M} = 0.7$ $M_{\odot}$ yr$^{-1}$ and 
$r_h = 1$ kpc for all flows. 
The four flows also have the same bubble mass 
$m_b = 10^5$ $M_{\odot}$ and heating factor $h = 6$, 
but the bubble drag heating efficiency has values 
$e_d = 0$, 0.1, 0.3 and 0.7.
Flows with progressively higher $T$, $T_b$ and $n$ near 
the heating radius $r_h$ correspond to increasing efficiencies.
By contrast, the circulation radii, 
velocities, filling factors and bubble radii are 
insensitive to the heating efficiency.

However, the flow with the largest heating efficiency,
$e_d = 0.7$ is less satisfactory because flow temperature
and particularly the apparent temperature 
$\langle T \rangle$ develop appreciable negative radial 
gradients near $r_h$.
Because of this rise in the flow temperatures, 
bubbles heated with the same $h$ rise to higher temperatures.
In addition, when $e_d = 0.7$ the density gradient develops an unstable 
positive density gradient within about 2 kpc. 
From these and other circulation flows with nonzero $e_d$ 
we conclude that the apparent temperature has no 
undesirable radial gradients as long as 
$e_d \lta 0.5$; this is a reasonable constraint in view of 
the uncertain nature of the bubble-flow heating process.

Figure 4 shows three flows assumed to be heated only by 
bubble expansion with efficiencies 
$e_{pdv} = 0.1$ 0.5 and 0.7 (and $e_d = 0$).
As before, the apparent emission-weighted temperature profiles 
(dotted lines in Fig. 4b) become strongly negative 
unless $e_{pdv} \lta 0.5$.

\section{Final Remarks}

Our primary objective in constructing these models is 
to explore the possibility that cooling flow gas 
can be heated at some small radius in such a manner 
that (1) no gas cools to very low temperatures and
(2) the apparent temperature gradient of the cooling 
flow does not become negative in the central regions. 
We have demonstrated that flows satisfying these 
two criteria are indeed possible. 
With appropriately chosen parameters, buoyant bubbles can 
carry the cooling flow mass ${\dot M}$ away from the 
core with velocities sufficiently large that 
the apparent (emission-weighted) 
temperature gradient $d\langle T \rangle/dr$
is virtually unaffected by the hot gas inside the bubbles. 
In addition, the local 
interbubble gas can be heated by the bubbles 
with up to 25 -- 50 percent efficiency 
without adversely disturbing the temperature gradient. 
This positive assessment of idealized 
circulation flows contrasts with the 
less satisfactory results of 
our numerical simulations in which all cooling 
flows that were sufficiently heated to noticeably reduce 
the central cooling ${\dot M}$ 
also resulted in strongly negative apparent temperature 
gradients 
(Brighenti \& Mathews 2002; 2003).

However, the bubble return model clearly fails if 
the bubbles are too small. 
As $m_b$ decreases, the large number of bubbles 
required to transport ${\dot M}$ 
(particularly near the central heating source at $r_h$)
consumes all the available volume, entirely displacing the 
cooling inflow (Condition 21). 
In these circumstances the heated central region would 
grow in size and would almost certainly result in 
an extended hot thermal core and $d\langle T \rangle/dr < 0$,  
which is not generally observed in X-rays. 
The circulation flows we consider do not specify 
the physical nature of the heating process 
interior to the heating radius $r_h$.
Consequently, if the bubbles are heated too slowly, 
there is some concern that emission from $r < r_h$
could result in negative apparent 
$d\langle T \rangle/dr$ and this would need to be 
considered in a more detailed model.
We have found that 
the problems of bubble congestion near $r_h$ 
and emission within $r_h$  
cannot be alleviated by increasing 
the radius $r_h$ at which heating occurs. 
Finally, bubbles of low mass also experience more drag, 
move more slowly and, if $m_b$ is too small, can fail to 
transport the mass arriving in the cooling inflow (Equation 22).

These difficulties with small bubbles 
clearly raise concerns about the possible fragmentation 
of larger bubbles. 
However, at the X-ray resolution of  
the observations we have so far, the bubble-like
disturbances have sizes $r_b/r \sim 0.05 - 0.3$ that 
are consistent with those required in our circulation flows. 
It is also noteworthy that the 
X-ray images showing bubbles and surface brightness 
fluctuations are rather similar on all scales, 
from elliptical galaxies to clusters like Perseus.
The azimuthally random character of bubble 
or density fluctuations at all radii in the hot gas 
also seems to be rather universal. 
The azimuthally random orientation of bubbles may be difficult 
to explain with numerical simulations 
in which energy is deposited by a fixed jet; 
the jet orientation would need to change rapidly 
due to precession or other similar black hole disturbances.
Finally, the X-ray data so far indicates that the strong 
(or complete) reduction in the cooling rate ${\dot M}$ 
occurs universally in all hot gas flows, whereas only 
a small (but extensively observed) fraction 
of these flows contain powerful jet-like radio sources. 

Circulation flows are also limited in spatial extent, 
and this limitation becomes particularly acute as the bubble 
mass decreases. 
In all the galaxy-scale circulation flows that we considered, 
the circulation radius $r_c$ is limited to $r \lta 25$ kpc 
which may be comparable to observed regions having 
a multi-temperature spectrum. 
For example in NGC 5044 Buote et al. (2003) find 
evidence of significant temperature fluctuations out 
to $\sim 30$ kpc. 
Although  
small bubbles are disadvantageous for a variety of reasons, 
there is no obvious 
mechanism that favors the production of larger bubbles or 
guarantees that larger bubbles do not fragment. 
This must be viewed as an essential shortcoming of cooling flows 
with bubble return and the notion that ${\dot M}$ is small
or zero because of AGN heating. 
While it is likely that the circulation radius $r_c$ 
can be increased beyond $\sim 25$ kpc 
if the bubbles are supported 
with cosmic rays or magnetic fields, such bubbles may be unable to 
carry the requisite ${\dot M}$ away from the heated core. 

In our circulation flows, the kinetic energy and momentum of the 
rising bubbles is derived 
ultimately from the gravitational field 
and some of this energy and momentum is transported 
to the surrounding gas.
The momentum exchange reduces the velocity of inflowing gas, 
and may be analogous to the entrainment of local gas in 
many of the numerical simulations of individual large bubbles 
reviewed in Section 2.
Bubbles can directly heat the surrounding gas if they 
thermally dissolve into the interbubble gas before
reaching the radius $\sim r_c$ of equal entropy.
While we do not consider such thermal mixing here, 
it would almost certainly heat the interbubble flow and 
result in unfavorable apparent temperature gradients.
Some limited heating of the interbubble gas by 
adiabatic bubbles -- 
either by drag or $PdV$ mechanical heating or by 
entropy mixing -- may account for the
enhanced non-gravitational entropy observed in galaxy groups
(e.g. Ponman et al. 1999). 
However, the efficiency ($e_d$ and $e_{pdv}$) of 
such distributed heating would need to be fine-tuned to 
be consistent with the flow of both 
mass and entropy to large 
radii as required in circulation flow models 
and to satisfy observational constraints on the 
apparent temperature profile $\langle T \rangle (r)$.

The idealized circulation flows we describe here 
could be extended to
include cooling inflow from $r > r_c$,
constant mass loss from stars, 
bubble fragmentation, 
cosmic ray pressure in the bubbles, 
a range of heating factors $h$ for the bubbles, 
bubble-flow thermalization, and
a distribution of bubble masses $m_b$ in each flow.
It seems likely however that the filling factor 
and mass flow constraints (Equations 21 and 22) 
must still be valid for any flow in which all or most 
of the mass flow ${\dot M}$ is carried outward by 
heated regions.
It is likely that our less complicated circulation flows
contain essential features that are generic to
all centrally heated flows 
designed to greatly reduce the cooling rate ${\dot M}$.

The idealized circulation flows we describe here 
could only apply for a limited time because of 
secular increase in the gas mass as 
new gas enters the circulating region. 
For example, if all the gas ejected from evolving red 
giant stars merges into the hot 
phase, the characteristic time for the hot gas density to 
increase appreciably in NGC 4472 is 
$t_* \sim \rho / \alpha_* \rho_* 
\approx 4.2~(r_c/10~{\rm kpc})^{1.18}$ Gyr where 
$\alpha_* = 4.7 \times 10^{-20}$ sec$^{-1}$ is the 
current specific rate of stellar mass loss. 
Moreover, in many galaxy/group scale cooling flows, including 
NGC 4472, the hot gas 
extends far beyond the circulation radii $r_c$ we calculate 
here. 
Therefore, the cooling inflow of gas crossing radius 
$r_c$ provides another secular increase in the 
mass of gas in the circulation region. 
As the mass of circulating gas slowly increases,
the circulation flow ${\dot M}$ is also expected to increase
and the amount of heating $h$ must decrease to maintain 
the circulation. 
Ultimately, a cooling event is expected (Equation 22). 
When the central ${\dot M}$ is strongly reduced 
in our numerical simulations of centrally heated cooling flows,
the flow cools episodically at larger radii in the flow 
(Brighenti \& Mathews 2002; 2003).
It remains to be determined if intermittent cooling events 
can agree with the X-ray observations, 
but the observations we have so far 
are consistent with no cooling at all.
Finally, the hot bubbles also carry
Type Ia supernovae (SNIa) iron enrichment out to $r_c$,
broadening the iron abundance profile relative to the stars, 
which agrees qualitatively with observations. 
The iron from SNIa acts like tracer particles that can 
reveal the direction and magnitude of the 
radial flow of hot gas
(Mathews \& Brighenti 2003).



\vskip.4in
Studies of the evolution of hot gas in elliptical galaxies
at UC Santa Cruz are supported by
NASA grants NAG 5-8409 \& ATP02-0122-0079 and NSF grants  
AST-9802994 \& AST-0098351 for which we are very grateful.



\clearpage
\vskip.1in
\figcaption[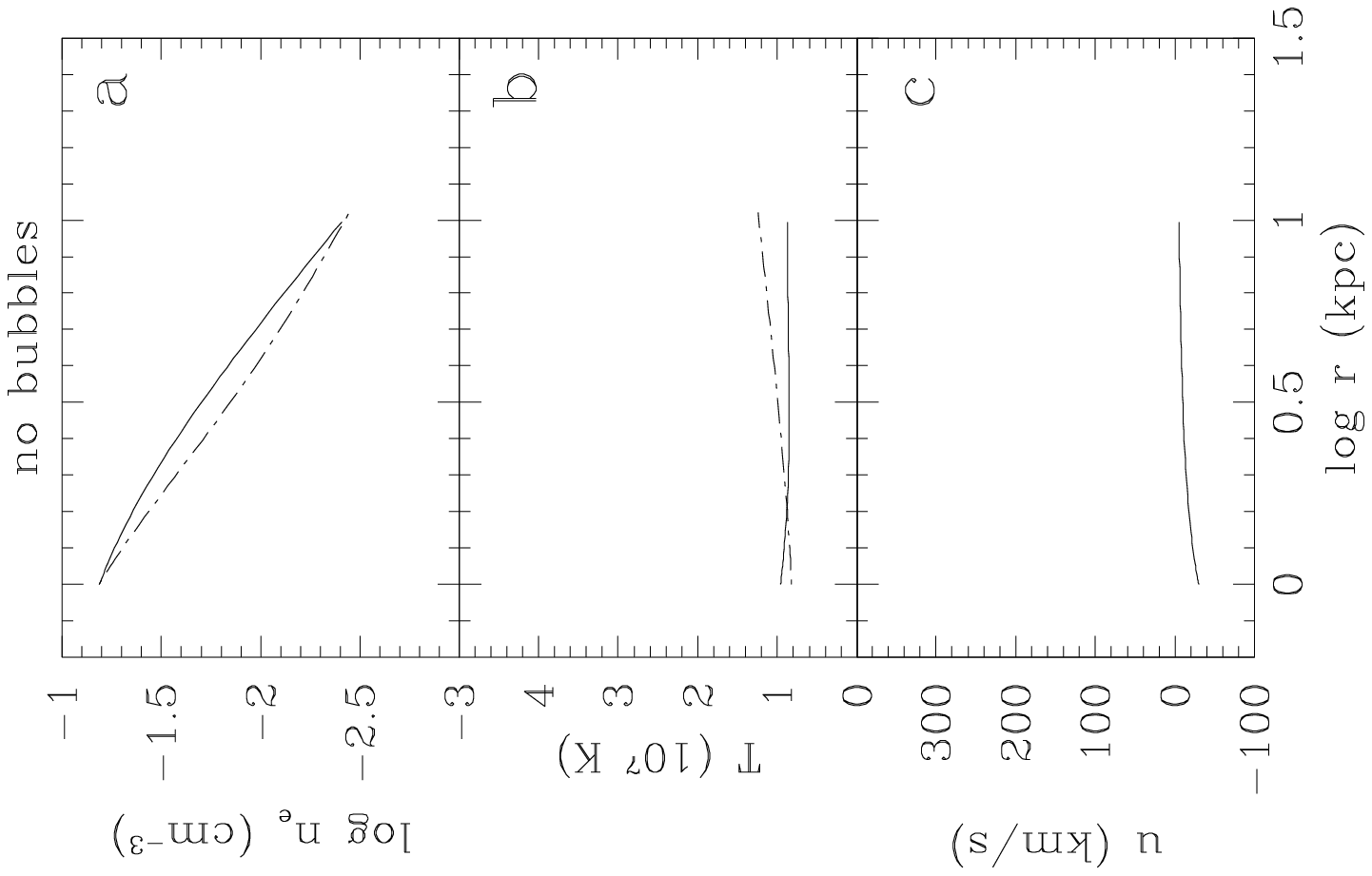]{
Panels a, b and c show respectively the gas density, 
temperature and flow velocity for a bubble-free cooling flow 
from $r_c = 10$ kpc to $r_h = 1$ kpc. 
These solutions of Equations (4) and (5) are designed so that 
the gas density matches the observed density in NGC 4472 
at $r_c$ and $r_h$. 
The observed temperature and density profiles for NGC 4472
are shown with dot-dashed lines.
\label{fig1}}

\vskip.1in
\figcaption[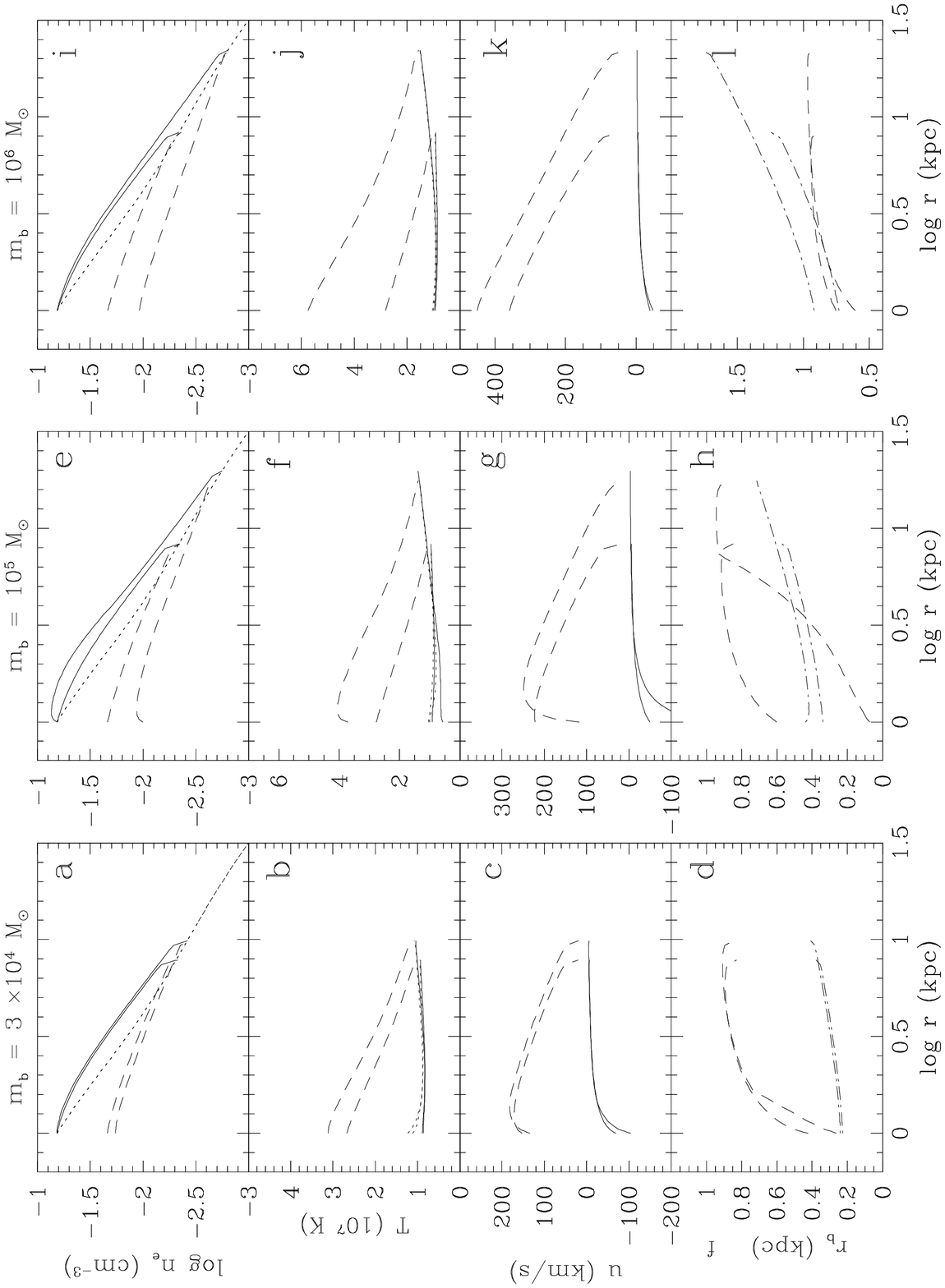]{
Six circulation flows with no heating from bubble-flow 
interactions ($e = 0$). 
Panels {\it a, b, c} and {\it d} refer to flows with bubbles of 
mass $m_b = 3 \times 10^4$ $M_{\odot}$.
Panels {\it a, b,} and {\it c} show respectively 
the gas density, temperature and velocity in the cooling inflow  
({\it solid lines}) and the gas density, 
temperature and velocity of the outward moving bubbles 
({\it dashed lines}). 
In panel {\it a} the observed 
density profile in NGC 4472 is shown with a {\it dotted line}.
Panel {\it b} shows the emission-weighted local temperature 
$\langle T \rangle$ of the flow plus bubbles ({\it dotted line}).
Panel {\it d} shows the volume filling factor 
$f(r)$ of the cooling inflow 
({\it dashed lines}) and the radius of the bubbles $r_b(r)$  
({\it dash-dotted lines}).
The pair of flows shown in panels {\it a - d} correspond to 
heating factors $h = 3$ and 3.6. 
Flows with larger $h$ extend to larger radii $r_c$.
The second column of panels ({\it e - h}) shows the same information 
for flows with bubbles of mass $m_b = 10^5$ $M_{\odot}$
with and $h = 3$ and 6.5.
Finally, the third column of panels ({\it i - l}) 
shows the same information
for flows with bubbles of mass $m_b = 10^6$ $M_{\odot}$
with $h = 3$ and 6.
\label{fig2}}

\vskip.1in
\figcaption[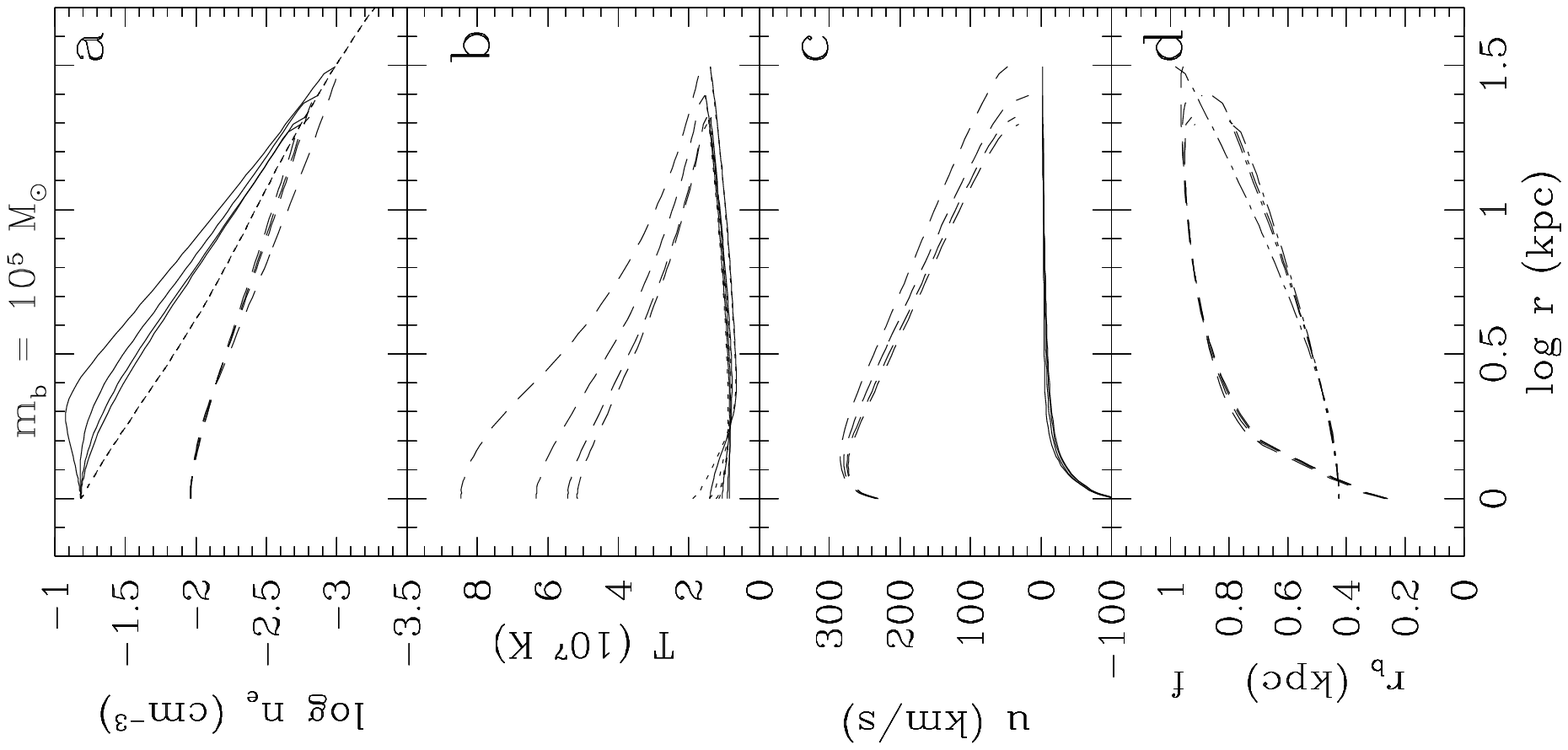]{
Four circulation flows with heating from bubble-flow
interactions at various drag efficencies $e_d$ 
(with $e_{pdv} = 0$).
The labeling scheme is the same as in Figure 2.
All solutions correspond to bubbles of mass 
$m_b = 10^5$ $M_{\odot}$ and heating factors $h = 6$.
The four solid curves in panel {\it a} show cooling inflows 
with progressively larger heating efficiencies,
$e = 0$, 0.1, 0.3 and 0.7.
The extent of the flow $r_c$ increases with 
the heating efficiency $e_d$, which allows 
each solution curve to be 
identified with the appropriate efficiency.
\label{fig3}}

\vskip.1in
\figcaption[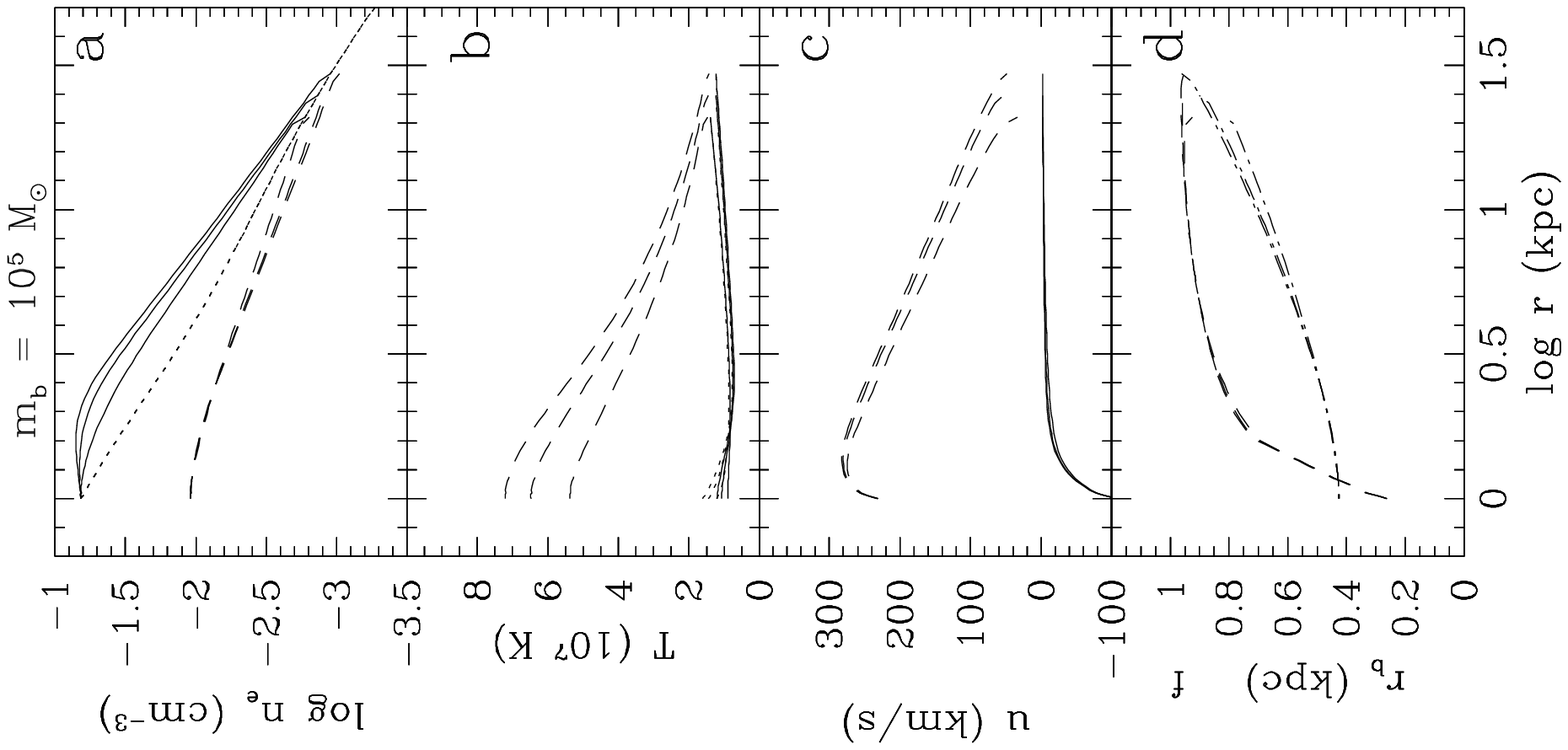]{
Three circulation flows heated by expanding bubbles
with various efficencies $e_{pdv}$
(with $e_d = 0$).
The labeling scheme is the same as in Figures 2.
All solutions correspond to bubbles of mass 
$m_b = 10^5$ $M_{\odot}$ and heating factors $h = 6$.
The three solid curves in panel {\it a} show cooling inflows
with progressively larger heating efficiencies,
$e_{pdv} = 0.1$, 0.5 and 0.7.
The extent of the flow $r_c$ increases with
the heating efficiency $e_d$, which allows
each solution curve to be
identified with the appropriate efficiency.
\label{fig4}}

\end{document}